\documentclass[final]{svjour2}
\usepackage{graphicx}
\usepackage{rotating}
\usepackage{amssymb}
\usepackage{mathptmx}
\usepackage[numbers]{natbib}
\makeatletter
\journalname{Journal of Low Temperature Physics}

\bibpunct{}{}{,}{s}{}{,}

\begin{document}

\newcommand{\hdblarrow}{H\makebox[0.9ex][l]{$\downdownarrows$}-}
\title{Macroscopic excitations in confined Bose-Einstein condensates, searching for quantum turbulence}

\author{R. Zamora-Zamora$^1$ \and O. Adame-Arana$^1$ \and V. Romero-Rochin $^1$}

\institute{1: Instituto de F\'{\i}sica, Universidad Nacional Aut\'onoma de M\'exico,\\ 09540 M\'exico, DF, Mexico USA\\
Tel.: +52 55 5622 5096\\ Fax: +52 55 5622 5015\\
\email{romero@fisica.unam.mx}}

\date{\today}

\maketitle


\begin{abstract}

We present a survey of macroscopic excitations of harmonically confined Bose-Einstein condensates (BEC), described by Gross-Pitaevskii (GP) equation, in search of routes to develop quantum turbulence. These excitations can all be created by phase imprinting techniques on an otherwise equilibrium Bose-Einstein condensate. We analyze two crossed vortices, two parallel anti-vortices, a vortex ring, a vortex with topological charge $Q = 2$, and a tangle of 4 vortices. Since GP equation is time-reversal invariant, we are careful to distinguish time intervals in which this symmetry is preserved and those in which rounding errors play a role. We find that the system tends to reach stationary states that may be widely classified as having either an array of vortices with collective excitations at different length scales or an agitated state composed mainly of Bogoliubov phonons. 

\end{abstract}

\section{Introduction}

Turbulence, be it classical or quantum, remains as a fundamental problem in the dynamics of fluids. Its high complexity makes it a very difficult, yet a fascinating state of matter in which one of the challenges is to describe it in simple terms. While there does not exist a clear definition or specification of what turbulence should be, there are several aspects or signatures that indicate whether or not one is facing a turbulent flow. Most notorious is the cascade of energy \cite{Kolmogorov} through vortices or eddies, in the classical case, or by a tangle of vortices \cite{Feynman} in the quantum version. Among the relevant questions one must deal with, specially in the quantum regime, is  the initiation or generation of such a complex state\cite{Parker_Adams_PRL_2005,Kobayashi_Tsubota_JLTP_2006,
Kobayashi_Tsubota_PRA_2007,BRDV_PRA_2014} and its characterization.
\cite{AC_White_JOP_CS_2011,AW_Baggaley_PRL_2012} 
These issues motivate the study presented in this article. We point out that there are already experimental realizations of quantum turbulence\cite{Henn-PRL2005,Henn-JLTP2010,Anderson-PRL2014} as well of studies of the route to turbulence.\cite{Kobayashi_Tsubota_PRA_2007,Tsubota-JLTP2008,Seman-LPL2011}

In this work, through the Gross-Pitaevskii (GP) model\cite{Gross,Pitaevskii} of an ultracold quantum gas, and assuming that arbitrary phase-imprinted states can be generated\cite{Williams_Holland_Nature_1999,
Matthews_Cornell_PRL_1999,Leanhardt_Ketterle_PRL_2002,Shibayama_JOPB_2011}  we present here a survey of macroscopic excitations that lead a Bose-Einstein condensate (BEC), confined in an external harmonic potential, to stationary agitated or chaotic states that, under appropriate conditions, may be considered to pass through turbulent transients or remain as such. Although the details of the model and of the different excitations that we explore are presented below, we want to advance here a general result that we find may be of relevance in a more complete study. This is the fact that the GP model is capable of showing a clear transfer of energy, initiated with the large scale imprinted excitation by external means, and ending up in an effective ``stationary" state, sometimes with a stable array of vortices with large scale collective excitations\cite{Stringari,Stringari_scissors}, and other times without traces of the initial excitation, but with the presence of Bogoliubov phonons.\cite{Bogoliubov} The former shows spectra close to Kolmogorov law $k^{-5/3}$, while the latter a spectral law with a positive exponent. We analyze the numerical solution of GP equation in terms of the spectra of the kinetic energy and  in terms of power spectra of the time evolution of the steady state. While we do not claim that such an analysis permits discerning the presence of turbulence, we find that those spectra do serve as signatures of the reached stationary states as well of the evolution towards it.

The mentioned decay into a stationary or steady state is not a true irreversible\cite{Tsubota_JLTP_2000} behavior of GP but actually a complicated non-linear dephasing effect, since GP obeys time-reversal invariance. The evolution appears as ``irreversible" but this is due to numerical rounding errors. We present a detailed study of such an effect in order to understand these stationary states. We believe it is relevant to take these aspects into account since dephasing contributions will compete with true irreversible development induced by actual dissipative effects - not considered in the present study. In this work, we limit ourselves to the numerical solutions of the time-reversal invariant GP equation, valid for temperatures as close as possible to absolute zero, but we recognize that in order to make a direct comparison with actual experiments, being these at finite temperature, one must take into account dissipative effects. There are already efforts along these lines \cite{Proment,Kobayashi_Tsubota_PRL_2005,ZNG,Jackson_PRA_2009,Reeves,Allen_PRA_2013,Allen_PRA_2014} including the effects of the thermal cloud that surrounds a confined BEC. True stationary states, of course, are found in those cases

\section{Gross-Pitaevskii quantum fluid}

Our model is summarized in the Gross-Pitaevskii equation\cite{Gross,Pitaevskii} for a three dimensional, one-component BEC confined by an external potential,
\begin{equation}
i \hbar \partial_t
\psi=-\frac{\hbar^{2}}{2m}\nabla^{2}\psi+V_{ext}\psi+g\left|\psi\right|^2\psi .
\label{gross_pit}
\end{equation}
The mean-field interaction coupling term $g = 4 \pi \hbar^2 N a/m$, where $N$ is the number of bosonic atoms of mass $m$, and $a$ is the $s$-wave scattering  length, assumed positive throughout. The GP wave function $\psi$ is normalized to one. Although our numerical analysis will be limited to an isotropic harmonic potential $V_{ext} = 1/2 m \omega \vec r^2$, there are several comments and aspects that should be generic for any potential. In particular, due to the problem at hand, we are interested in also posing relevant questions in terms of a hydrodynamic formulation\cite{Gross,Nore,Fetter_JLTP_2010}. This is done as follows.

Using the transformation $\psi = \sqrt{\rho} e^{i\phi}$ and the identification $\vec v = (\hbar/m)\nabla \phi$, where $\rho$ is the particle density and $\vec v$ may be interpreted as the fluid velocity field, one finds the following hydrodynamic equations,
\begin{equation}
 \partial_t \rho + \nabla \cdot \left(  \rho\vec v \right)
= 0 ,\label{nfield}
\end{equation}
and
\begin{equation}
\rho \partial_t \vec v  + \rho \left(\vec v  \cdot \nabla\right) \vec
v=\nabla \cdot \tilde \sigma -\frac{\rho}{m}\nabla V_{ext}.
\label{vfield}
\end{equation}
where the stress tensor is,
\begin{equation}
 \sigma_{ij}= \left(\frac{\hbar^{2}}{4m^{2}}\partial_{kk}
\rho -\frac{g}{2m}\rho^{2}\right)\delta_{ij}-
\frac{\hbar^{2}}{m^{2}}\partial_{i}
\sqrt{\rho}\partial_{j}\sqrt{\rho} \label{presiones}
\end{equation}
and with the additional condition that the fluid is irrotational $\nabla \times \vec v = 0$. It is of interest to note that the stress tensor has a hydrostatic pressure $p_0=g \rho^2/2m$, due to the interatomic interactions, and other volumetric and shear stresses of pure quantum origin, namely, all those proportional to $\hbar^2$. The above hydrodynamic equations have been also previously derived by Grant.\cite{Grant}

As it is known, and we discuss and review it here, there is a ground stationary state, $\psi(\vec r,t) = e^{-i\mu t/\hbar} \phi_0(\vec r)$, with $\mu$ the chemical potential and $\phi_0(\vec r)$ the ground state macroscopic wave function, solution to,
\begin{equation}
\mu 
\phi_0=-\frac{\hbar^{2}}{2m}\nabla^{2}\phi_0+V_{ext}\phi_0+g\left|\phi_0\right|^2\phi_0 .
\label{gross_pit-0}
\end{equation} 
For $V_{ext}$, $g$ and $m$ given, the chemical potential $\mu$ and $\phi_0$ are uniquely given.  The wave function $\phi_0(\vec r)$ can be accurately found numerically.\cite{Zeng,RAZA} This is our starting point for all the forthcoming calculations. From the hydrodynamic point of view, this fluid has an inhomogenous density $\rho_0(\vec r) = |\phi_0(\vec r)|^2$ and zero velocity field everywhere, $\vec v = 0$. Any macroscopic excitation on top of the state $\phi_0$ will necessarily evolve in time (except perhaps a line vortex at an axis, but this is unstable\cite{Castin}).

While there may be many more excitations, we consider for the moment three of them, vortices, collective modes, such as breathing or scissors modes,\cite{Stringari,Stringari_scissors} and sound waves, or Bogoliubov phonons,\cite{Bogoliubov} and their interactions. The line vortex, for $V_{ext} = 0$, is the actual solution given by Gross\cite{Gross} and Pitaevskii\cite{Pitaevskii}, with zero density at the vortex and the velocity field yielding a non-zero circulation,
\begin{equation}
 \oint \vec v \cdot d \vec l = \frac{2 \pi \hbar}{m} Q ,
 \end{equation}
with $Q = 1,2,\dots$ the topological charge. On the other extreme, we have that any small perturbation is composed of Bogoliubov sound waves, as we now briefly review.

Let us assume a stationary solution to the equations, which is a uniform one $\rho_0 = \mu/g$ if $V_{ext} = 0$, and the non-uniform solution $\rho_0(\vec r) = |\phi_0(\vec r)|^2$ for the harmonic trap, discussed above. In both cases, the velocity field is zero everywhere, $\vec v = 0$. Let us now  consider a perturbation in the free-field case, $V_{ext} = 0$,
\begin{equation}
\rho \approx \rho_0 + \rho_1(\vec r,t) \>\>\>{\rm and}\>\>\> \vec v \approx \vec v_1(\vec r,t) .
\end{equation}
Then, a simple linearization of the hydrodynamic equations, eqs. (\ref{nfield}) and (\ref{vfield}), yields,
\begin{equation}
\partial_{tt} \rho_1
-\frac{\hbar^{2}}{4m^{2}}\nabla^{2}\left(\nabla^{2}
\rho_1 \right)+\frac{g\rho_0}{m}\nabla^{2} \rho_1 =0 . \label{GP_lineal}
\end{equation}
This is a fourth-order wave equation whose solution is $\rho_1 = R_k e^{i (\vec k \cdot \vec r - \omega_k t)}$, with $R_k$ an arbitrary complex amplitude, yielding,
\begin{equation}
 \omega_k^{2}=\frac{\hbar^{2}k^{4}}{4m^{2}}+\frac{g\rho_0k^{2}}{m} \label{Bogo}
\end{equation}
which, as expected, is Bogoliubov dispersion relation for the elementary excitations in a weakly-interacting Bose gas \cite{Bogoliubov}. The velocity field is, in turn, found as $\vec v_1 = \rho_0 \vec k (R_k \omega_k/k^2) e^{i (\vec k \cdot \vec r - \omega t)}$, which shows that the waves (phonons) are longitudinal, also as expected. 

The third important excitation are those that are referred as collective, namely, excitations that involve the whole cloud, and that occur from such a length scale down to perhaps the vortex size, such as breathing or scissors modes. These can be studied in the Thomas-Fermi limit, namely, neglecting the kinetic energy in GP equation. A clear exposition of these modes may be found in the works of Stringari and coworkers.\cite{Stringari,Stringari_scissors} Suffice to say here that these are not uncoupled to the other excitations, vortices and phonons, as we will exemplify below.

\section{Macroscopic excitations, stationary states, and their energy spectra}

 In this section we study the time evolution of several initial states, described in detail below. We analyze their decay to their respective ``stationary" states as well as the time evolution of frequency and wavenumber spectra of the different energy components of the motion. The different initial states are prepared, first, by allowing the system achieve its ground state $\phi_0(\vec r)$ (for an isotropic harmonic potential of frequency $\omega$) and then by imprinting different types of vortex perturbations. In a typical case,  the initial state is
 \begin{equation}
 \psi(\vec r, 0^+) = \phi_0(\vec r) \> e^{i Q \theta(\vec r)}, \label{excite}
 \end{equation}
 where the imposed phase functions $\theta(\vec r)$ are equivalent to the generation of a velocity field representing a given vortex excitation. $Q$ is the vortex topological charge. We study the crossing of two orthogonal vortices (3.1);  the collision of two anti-vortices (3.2); an off-center vortex ring (3.3); a vortex of charge $Q = 2$ (3.4); and a tangle of 4 vortices (3.5). In the parenthesis we have indicated the section below in which we analyze them.
 
Our calculations are performed using parallel computing with Graphic Processors Units (GPU), which allows us to develop large and fast calculations; details of the numerical methods and programming will be reported elsewhere. We use a spatial grid of size $256^3$, in double-precision and with time steps of $\Delta t = 0.0005$, in dimensionless units $\hbar = m = \omega = 1$, and with a coupling $g = 8000$. This corresponds fairly well to a gas of $N = 1.4 \times 10^5$ atoms of $^{87}$Rb in an isotropic trap of frequency $\omega = 2 \pi (100)$ Hz with a scattering length $a = 100$ \AA, and with a dimensionless equilibrium chemical potential of $\mu =$ 19.63. The unit of time corresponds to 0.00159 seconds.

While we monitor wave function normalization, energy and angular momentum conservation in order to partially ensure numerical convergence of our calculations, we have also used a further dynamical criterion since we are interested in time evolution behavior rather than in static properties. This criterion arises from the observation that the prepared initial states all do seem to ``relax" to ``stationary" states. This is an important aspect to analyze since the lack of dissipation in GP equation should prevent the system from truly relaxing to a thermal equilibrium state. In the present case, given that GP is time-reversal invariant, one faces a {\it dephasing} mechanism rather than a dissipative relaxing one. This is the more important for non-linear equations solved by numerical means which, because of their concomitant round-off errors, very commonly give rise to a loss of the intrinsic reversibility of the dynamical equation, thus yielding an apparent irreversible behavior.
 
 To be more precise, GP as well as Schr\"odinger equation, are time-reversal invariant under the transformation $t \to -t$ and $\psi \to \psi^*$. Hence, if at any time $t$ we change $\psi$ to $\psi^*$, the system returns to its initial state at a time $2t$. On the one hand, one expects that if time $t$ is very large, due to rounding errors, eventually the state will not return to its original one. That is, for long times we can no longer assert that the observed evolution does correspond to the original initial state. On the other hand, however, due to the non-linearity of the equation, and aided by numerical errors, the state of the system can reach what one may call ``basins of attractions",\cite{Strogatz} such that, once the state enters one of those, it does not leave them anymore. We call these states {\it stationary}. We further find that, once the system reaches one of these states, time-reversal invariance is numerically restored. This is just an indication that those states are very stable, even against round off errors. This study is performed as follows.
 
\begin{figure*}[ht]
\begin{center}
\includegraphics[width=1.0\textwidth]{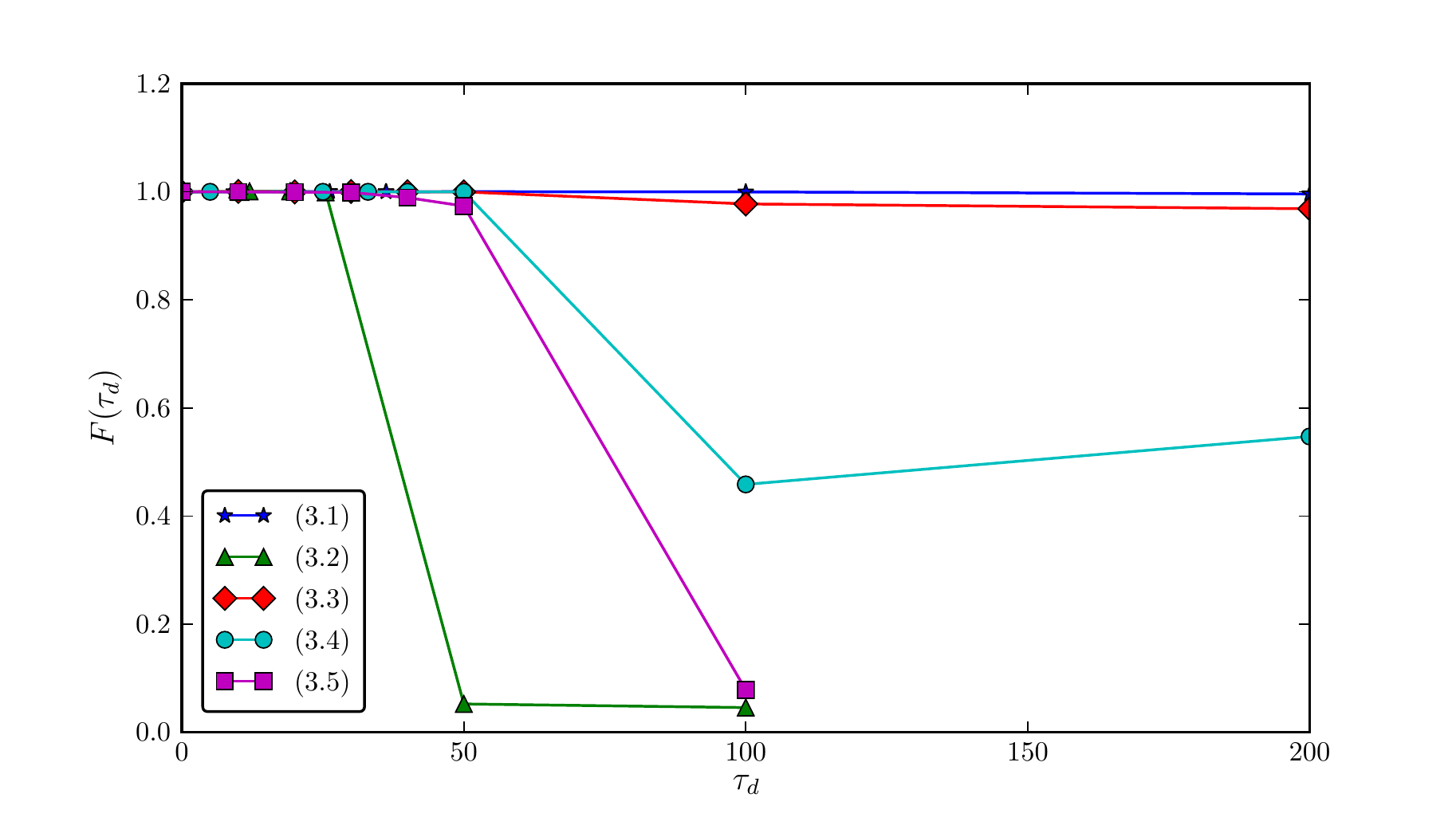}
\end{center}
\caption{(Color online) Fidelity $F(\tau_d)$ vs time $\tau_d$, see Eq. (\ref{F0}). The labels refer to the sections where the cases are analyzed, (3.1) crossing of two orthogonal vortices; (3.2) collision of two anti-vortices; (3.3) off-center vortex ring; (3.4) vortex of charge $Q = 2$; (3.5) a tangle of 4 vortices.}
\label{fidelity0All}
\end{figure*}

Let $\psi(\vec r,0)$ be any of the initial states at time $t = 0$, and let $\psi(\vec r, t)$ its GP solution at time $t$. At time $t = \tau_d$ we make the transformation $\tilde \psi(\vec r, t) = \psi^*(\vec r, t)$, and evolve $\tilde \psi(\vec r, t)$ for another interval of time $\tau_d$. Time reversal invariance demands that $\tilde \psi(\vec r, 2 \tau_d) = \psi^*(\vec r,0)$. Thus, for different times $\tau_d$ we calculate the so-called fidelity
\begin{equation}
F(\tau_d) \equiv \int \tilde \psi(\vec r,2 \tau_d) \psi(\vec r,0) \> d^3 r .\label{F0}
\end{equation}
If there were no loss of time reversal invariance, $F(\tau_d) = 1$ for all times $\tau_d$. In Fig. \ref{fidelity0All} we show such a calculation for all the cases we study for up to time $\tau_d = 200$. We see that case (3.1) is the most stable, followed by (3.3), while (3.2), (3.4) and (3.5) loose their time-reversal invariance for times longer than $\tau_d \approx 50$. Interestingly, all relax to a clear stable stationary state, as we now show. However, while in the first two cases one can affirm that the stationary state corresponds to the initial state, for the latter three we can only say that the initial state is {\it close} to the basin of attraction of the reached stationary state.

\begin{figure*}[ht]
\begin{center}
\includegraphics[width=1.0\textwidth]{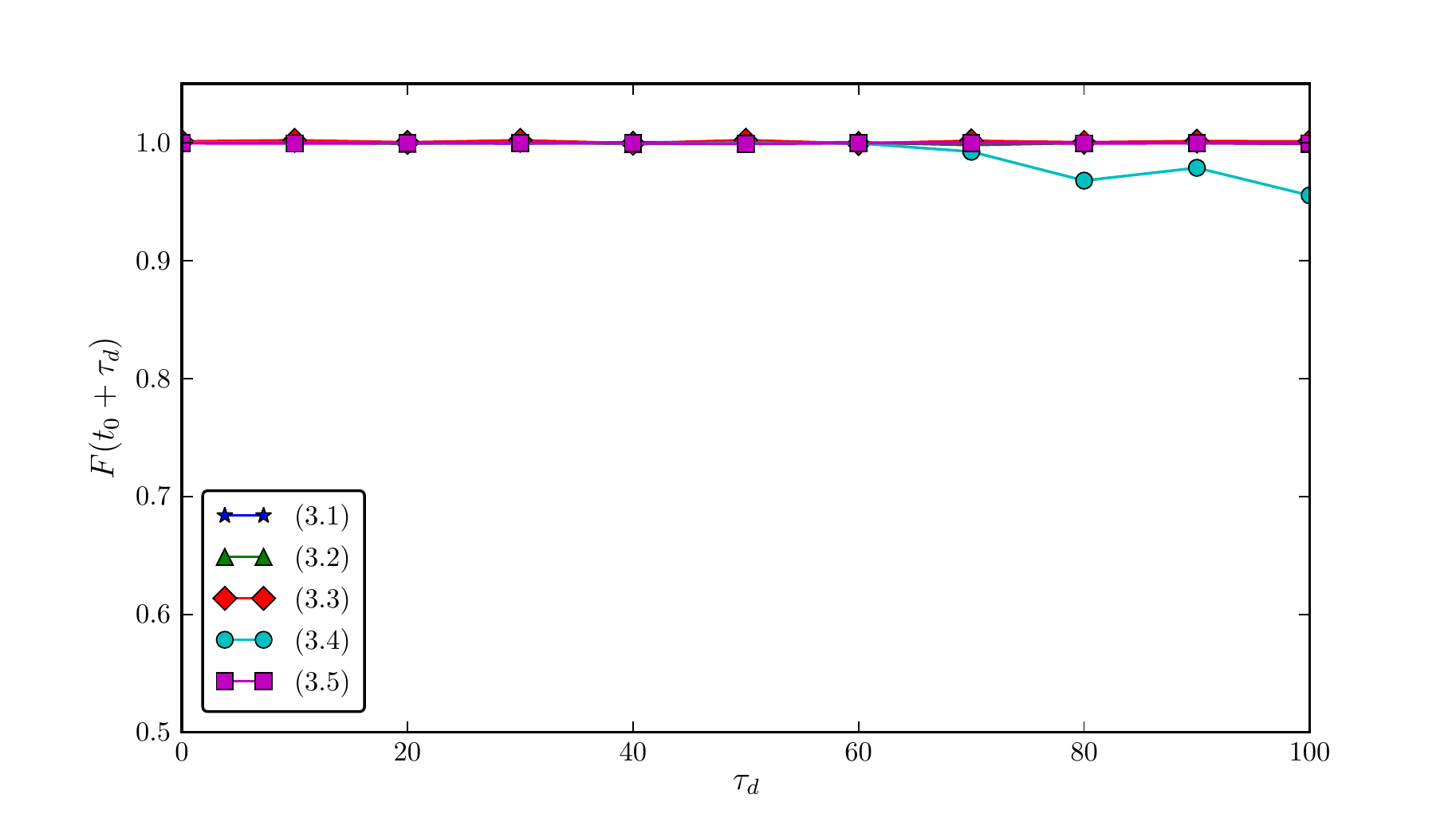}
\end{center}
\caption{(Color online) Fidelity $F(t_0+\tau_d)$ vs time $\tau_d$, where $t _0 = 200$,  see Eq. (\ref{Fs}).}
\label{fidelity200All_v1}
\end{figure*}

In order to verify the stability of the stationary state, we first let the system achieve it. Those states are typically reached by time $t = 200$. Therefore, we now calculate the fidelity at times $(t_0 + 2 \tau_d)$ with $t_0 = 200$ as the initial or reference state, that is,
\begin{equation}
F(t_0 + \tau_d) \equiv \int \tilde \psi(\vec r,t_0 + 2 \tau_d) \psi(\vec r,t_0) \> d^3 r .\label{Fs}
\end{equation}
 The result is shown in Fig. \ref{fidelity200All_v1}. With the exception of (3.4) that looses a bit of fidelity, the rest appear very stable.
 
As part of the statistical analysis of both the time evolution and the stationary state properties, we present the spectra of the incompressible part of the kinetic energy.\cite{Nore} The spectra is calculated as
\begin{equation}
K_{inc} = \int \> \frac{1}{2} \left( \rho \vec v^2 \right)_{inc} \> d^3r = \int \> \epsilon_\alpha (k) \> dk .\label{K}
\end{equation}
Of particular interest is the search for situations whether the incompressible spectra shows a Kolmogorov law $k^{-5/3}$ or not. 

In addition to the time evolution of the spectra, and in order to obtain a better understanding of the stationary state, we calculate the time evolution and the frequency spectrum of the modulus of the overlap between the state at the time $t_0 = 200$ and later times $t_0 + \tau$. This is
\begin{equation}
C(t_0+\tau) = \left| \int \psi*(\vec r, t_0 + \tau) \psi(\vec r,t_0) d^3 r\right| .\label{overtime}
\end{equation}
We denote its Fourier transform as $\tilde C(\omega) = {\cal F} \left[C(t_0+\tau)\right]$. As we shall see, this function samples the contribution of different excitations in the stationary state. This information can be correlated with the spectra $\epsilon_{inc}(k)$ to elucidate whether the excitations are vortices, collective modes and/or phonons.

 \subsection{Crossing of two orthogonal vortices.}

\begin{figure*}[ht]
\begin{center}
\includegraphics[width=1.0\textwidth]{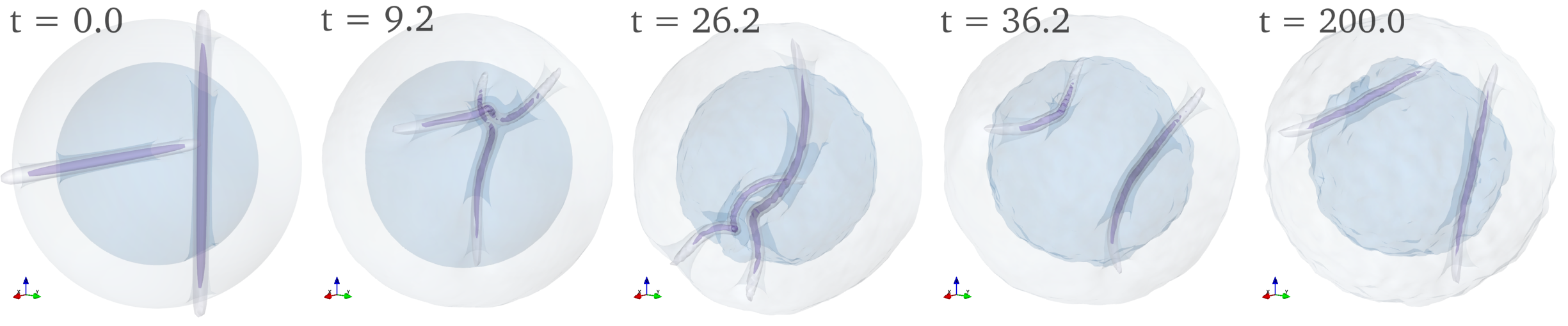}
\end{center}
\caption{(Color online) Snaphots of the magnitude of the velocity field for the collision of two orthogonal vortices.}
\label{cruzados}
\end{figure*}

In this case, the initial state is given by two line vortices at orthogonal directions, see Fig. \ref{cruzados}, one with $Q = +1$ parallel to the $x$-axis at $y = -2.0$ and $z = 0$, the other with $Q = -1$ parallel to the $z-$ axis at $x = 0$ and $y = 2.0$,
\begin{equation}
\theta(\vec{r}) = \arctan\left(\frac{y+2.0}{z}\right)-\arctan\left(\frac{x}{y-2.0}\right)
\end{equation}
As one may expect, see the set of snapshots for the magnitude of the velocity field in Fig. \ref{cruzados}, the vortices join and reconnect few times. We have observed three of these reconnections within a time interval of 50 units of time. Then, the vortices stop crossing and become almost but not quite parallel and keep orbiting around each other, in a stationary state. This late state shows, on top of the orbiting vortices, an agitated fluid that appears to be a superposition of collective modes at different lenght scales, with the presence of few phonons. Very interestingly, its incompressible energy espectra shows a scaling law quite close to Kolmogorov law. This is the most stable case since it reaches its stationary state preserving its time-reversal invariance, showing a stable fidelity times much longer than 200 units of time.

\begin{figure*}[ht]
\begin{center}
\includegraphics[width=1.0\textwidth,keepaspectratio]{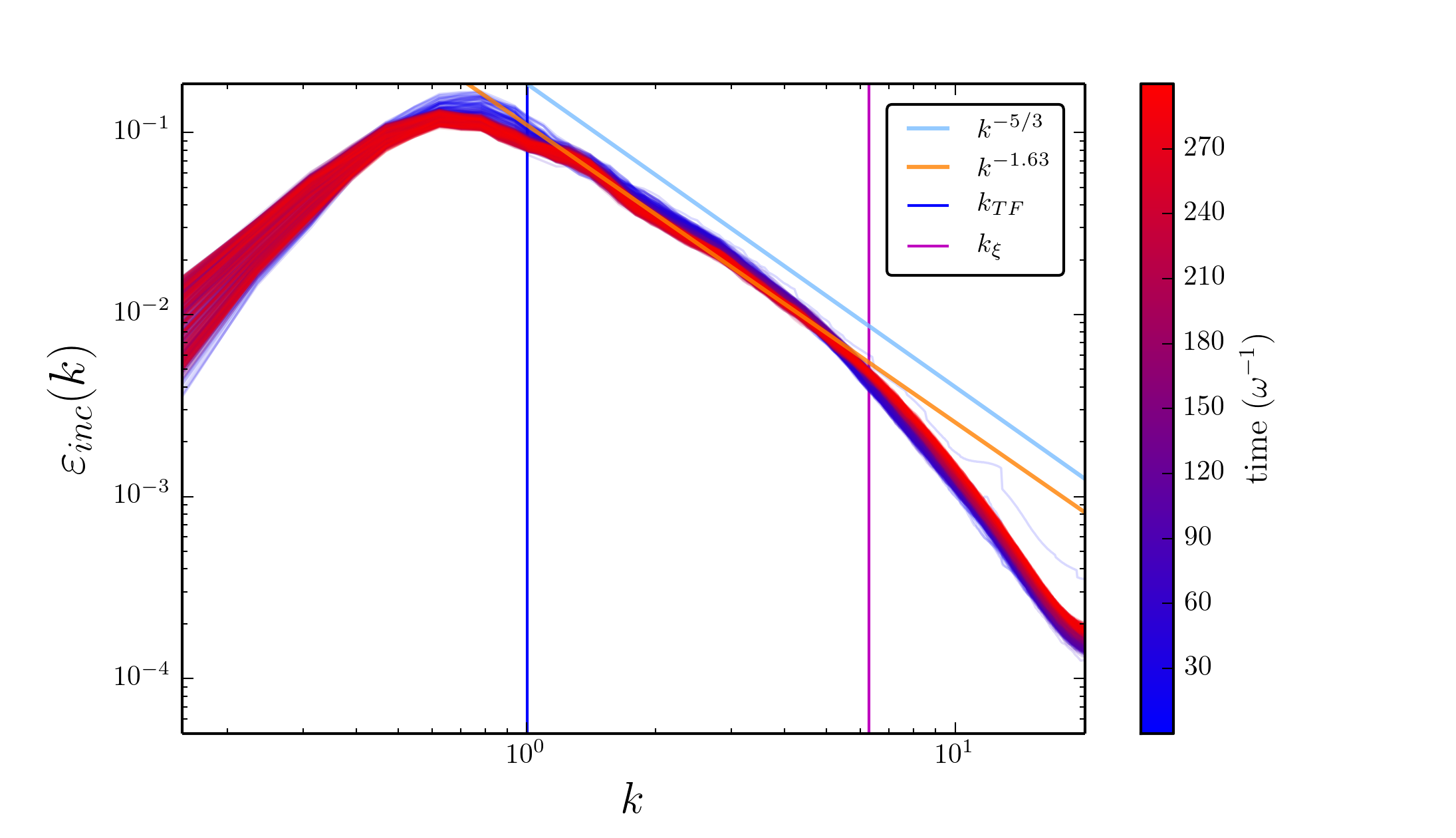}
\end{center}
\caption{(Color online) Time evolution of the incompressible kinetic energy spectrum of two orthogonal vortices. For reference, we mark the lengths $k_{TF} = 2 \pi/R_{TF}$ and $k_\xi = 2 \pi/\xi$, with $R_{TF}$  the Thomas-Fermi radius, the size of the cloud, and $\xi$ the vortex core size. We also show Kolmogorov slope $-5/3$ as well as the best fit for all slopes between $k_{TF} \approx 1.00$ and $k_\xi \approx 6.28$, equal to $- 1.63$.}
\label{ene-cruzados}
\end{figure*}

Fig. \ref{ene-cruzados} shows the time evolution of the incompressible kinetic energy spectrum. The compressible part is not shown being one order of magnitude smaller. The spectrum does not show complicated features in its evolution, varying very little in time and showing that although the initial state is somewhat unstable, it does not differ much from the stationary one.  One sees that the spectra is quite parallel to the $k^{-5/3}$ line in the segment of the spectrum between the Thomas-Fermi and the vortex size wavenumbers, $k_{TF}$ and $k_\xi$. The average slope is $-1.63$, very close to $-5/3$. As discussed further below, this appears as a signature of an algebraic cascade of energy between the vortices and collective excitations, with very little presence of phonons. 

\begin{figure*}[ht]
\begin{center}
\includegraphics[width=1.0\textwidth,keepaspectratio]{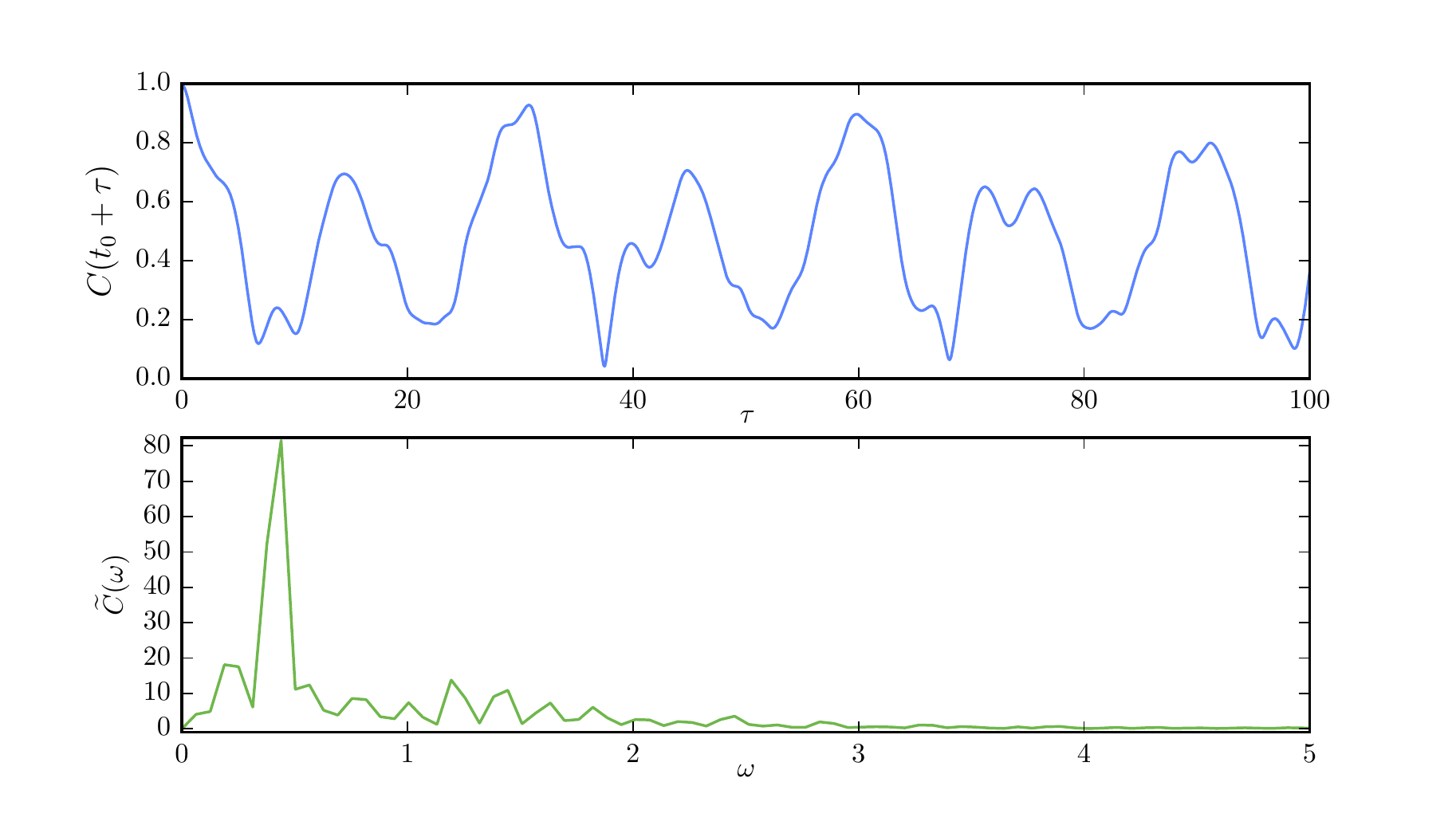}
\end{center}
\caption{(Color online) Overlap $C(t_0 + \tau)$ vs $\tau$, see Eq.(\ref{overtime}), for $t_0 =200$ in the upper panel, and its Fourier transform $\tilde C(\omega)$ vs $\omega$  in the lower one, for the initial state of two orthogonal vortices.}
\label{overlapCross}
\end{figure*}

The above conclusion may be further understood from the overlap evolution $C(t_0+\tau)$ and its Fourier transform, see Eq. (\ref{overtime}), as shown in Fig. \ref{overlapCross}. The largest peak corresponds to rotation frequency of the two vortices ($\omega \approx 0.42$), while the other frequencies are collective modes. The peaks in $\omega = 1$ correspond also to motion of the whole cloud in the harmonic potential; we have verified this by studying the center of mass motion of the cloud (not shown here). As we shall verify below, there seems not be an important contribution of phonons in this case.

\subsection{Collision of two anti-vortices}

\begin{figure*}[ht]
\begin{center}
\includegraphics[width=1.0\textwidth]{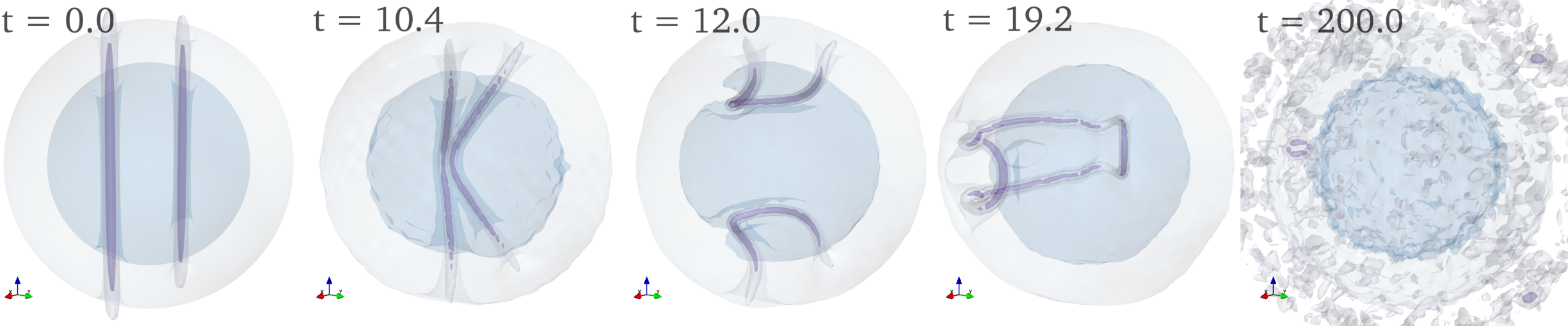}
\end{center}
\caption{(Color online) Snaphots of the magnitude of the velocity field for the collision of two anti-vortices.}
\label{collision}
\end{figure*}

Fig. \ref{collision} shows the evolution of the collision of two parallel vortices, along the $z$-axis, one with topological charge $Q = 1$, at $y = 2.0$, and the other with $Q = -1$ at $y = -2.0$, namely,
\begin{equation}
\theta(\vec{r}) = \arctan\left(\frac{x}{y-2.0}\right)-\arctan\left(\frac{x}{y+2.0}\right)
\end{equation}
The behavior of this case is quite the opposite of the two orthogonal vortices.
Referring to Fig. \ref{collision}, first, the vortices approach each other and reconnect ($t = 
10.4$) forming two vortices at directions orthogonal to the original ones ($t = 12.0$), but because 
they have opposite circulations, eventually join at their extrema forming a twisted vortex ring 
($t = 19.2$) that further folds into itself. Then, it appears to break into smaller rings that 
eventually become what appears to be a very agitated state with excitations and phonons at all 
length scales and in all directions. While this case seems similar to the Crow instability in 
uniform systems,\cite{Berloff-jphysA-2001,Simula-PRA-2011} the presence of the trap may affect its 
evolution. This case has the shortest dephasing time, see Fig. \ref{fidelity0All}, 
yet, once it enters the stationary state remains there, see Fig. \ref{fidelity200All_v1}.

\begin{figure*}[ht]
\begin{center}
\includegraphics[width=1.0\textwidth]{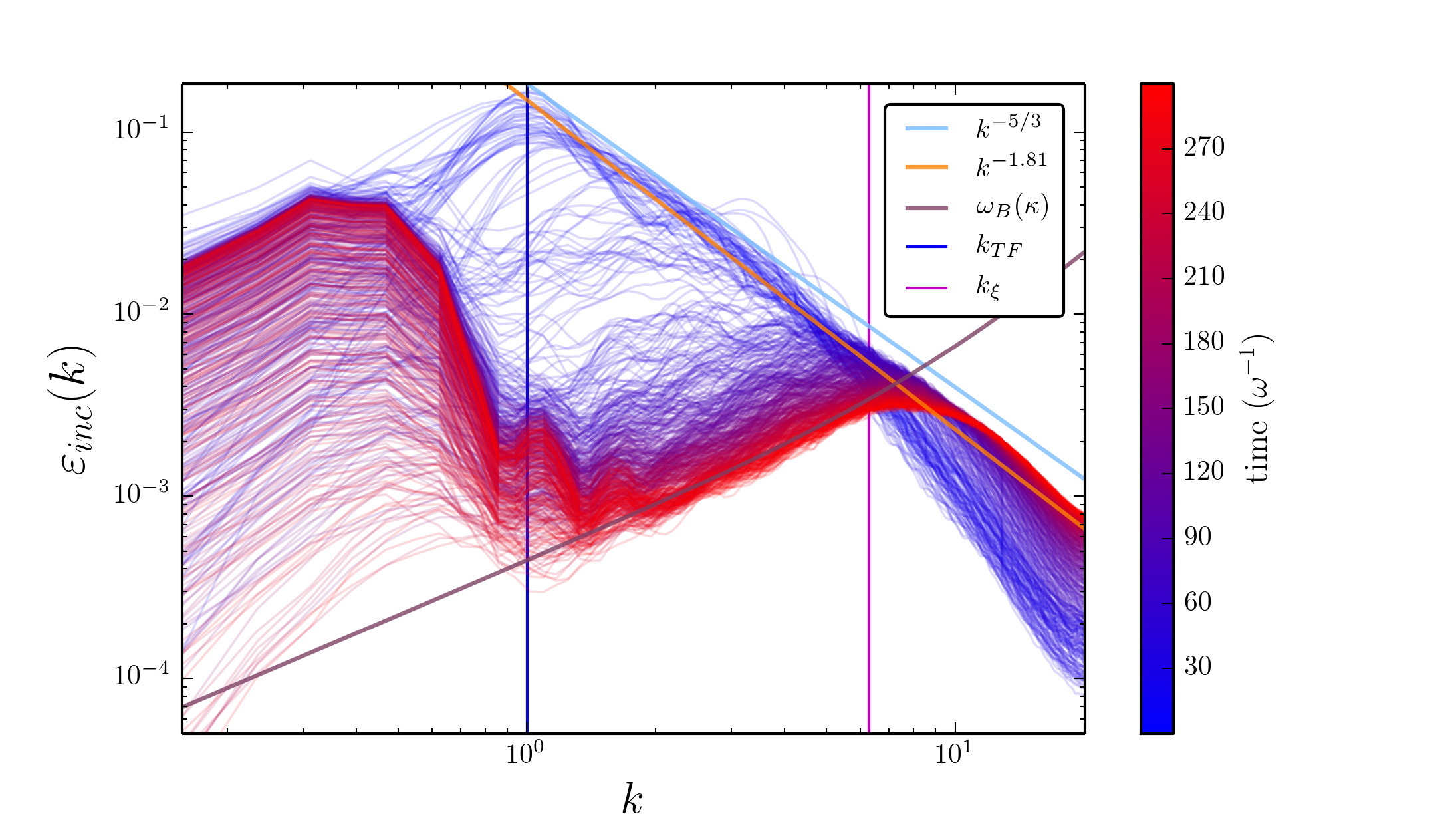}
\end{center}
\caption{(Color online) Time evolution of the incompressible kinetic energy spectrum of two anti-vortices. We show the length scales $k_{TF}$ and $k_\xi$, see Fig. \ref{ene-cruzados}. We also mark the slope $-5/3$ and the best slope fit of the first 60 units of time. The line marked as $\omega_B(k)$ is the Bogoliubov phonon dispersion relation given by Eq. (\ref{Bogo}).}
\label{ene-collision}
\end{figure*}

Fig. \ref{ene-collision} shows the time evolution of the energy spectrum of the collision of two anti-vortices. The figure shows that the system quickly evolves towards a stationary state, which for late times $t > 150$, is composed by a collective oscillation of the cloud, with the first peak about $k_{TF}$ followed by phonons excitations. The latter can be verified by the line indicated by $\omega_B(k)$ which is the dispersion relation of Bogolubov phonons, see Eq. (\ref{Bogo}); that is, the kinetic energy spectra is essentially composed of phonons in this case. The {\it compressible} contribution of the kinetic energy (not shown here) is quite similar to the incompressible one of Fig. \ref{ene-collision}, corroborating that the excitations are compressible, namely, acoustic excitations. 

\begin{figure*}[ht]
\begin{center}
\includegraphics[width=1.0\textwidth,keepaspectratio]{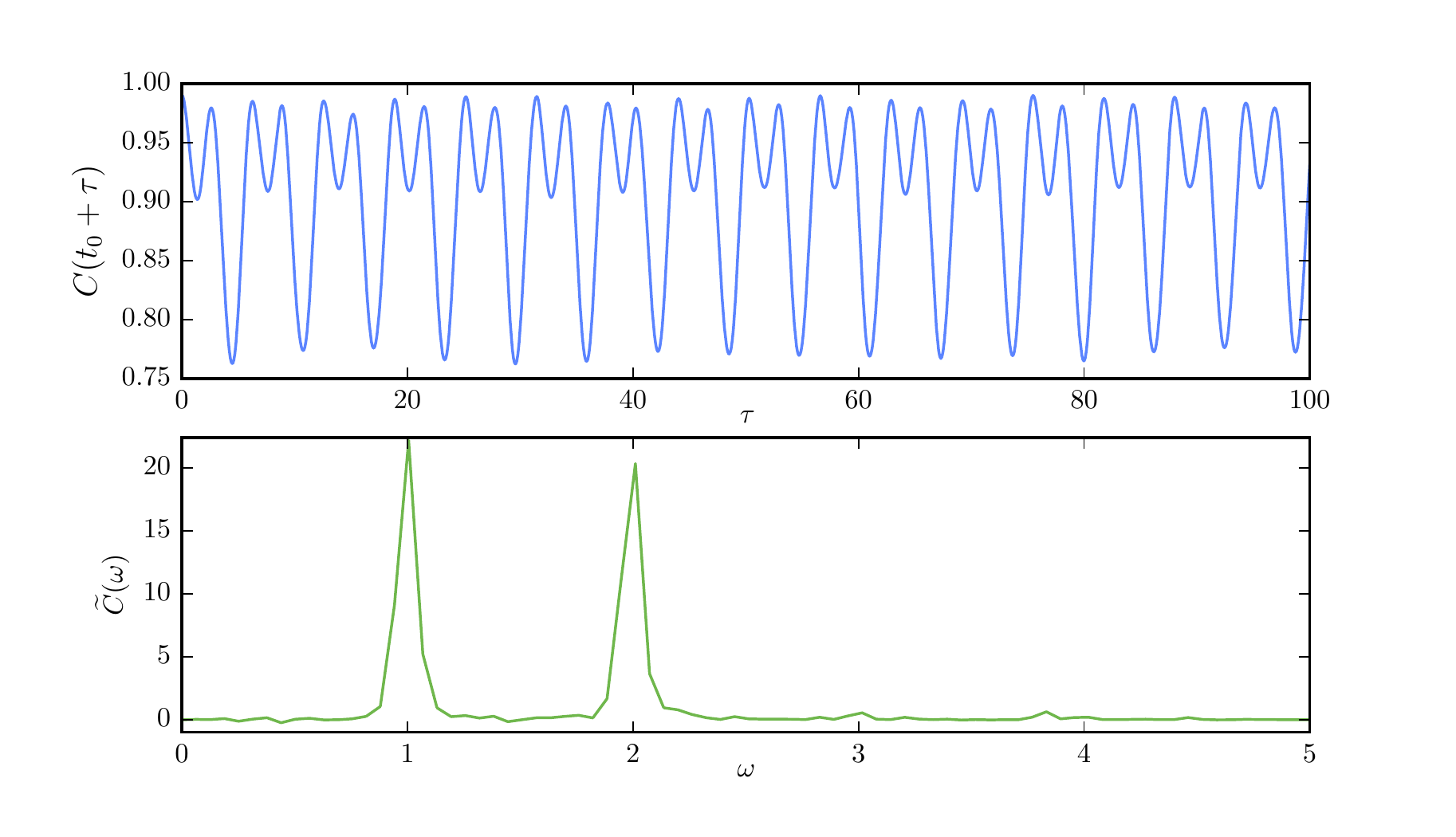}
\end{center}
\caption{(Color online) Overlap $C(t_0 + \tau)$ vs $\tau$, see Eq.(\ref{overtime}), for $t_0 =200$ in the upper panel, and its Fourier transform $\tilde C(\omega)$ vs $\omega$  in the lower one, for the collision of two anti-vortices.}
\label{overlapColl}
\end{figure*}

The overlap function $C(t_0 + \tau)$ in the stationary state, Fig. \ref{overlapColl}, shows no presence of collective excitations, except for two peaks at $\omega = 1$ and $\omega = 2$, which are motions of the center of mass, as directly verified from the numerical solution. 
 
\subsection{An off-center vortex ring}
 
\begin{figure*}[ht]
\begin{center}
\includegraphics[width=1.0\textwidth]{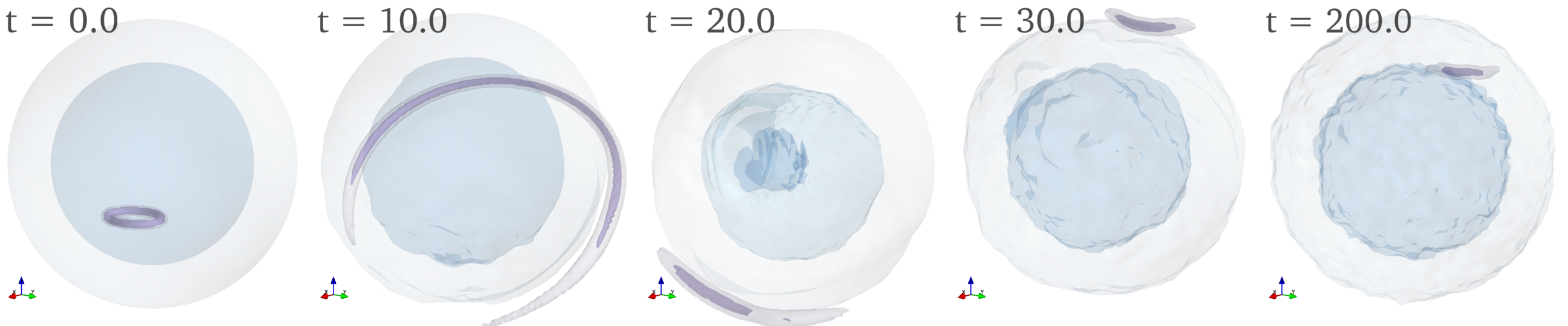}
\end{center}
\caption{(Color online) Snaphots of the magnitude of the velocity field for the evolution of a ring vortex (RingNC)}
\label{ring}
\end{figure*}

\begin{figure*}[ht]
\begin{center}
\includegraphics[width=1.0\textwidth]{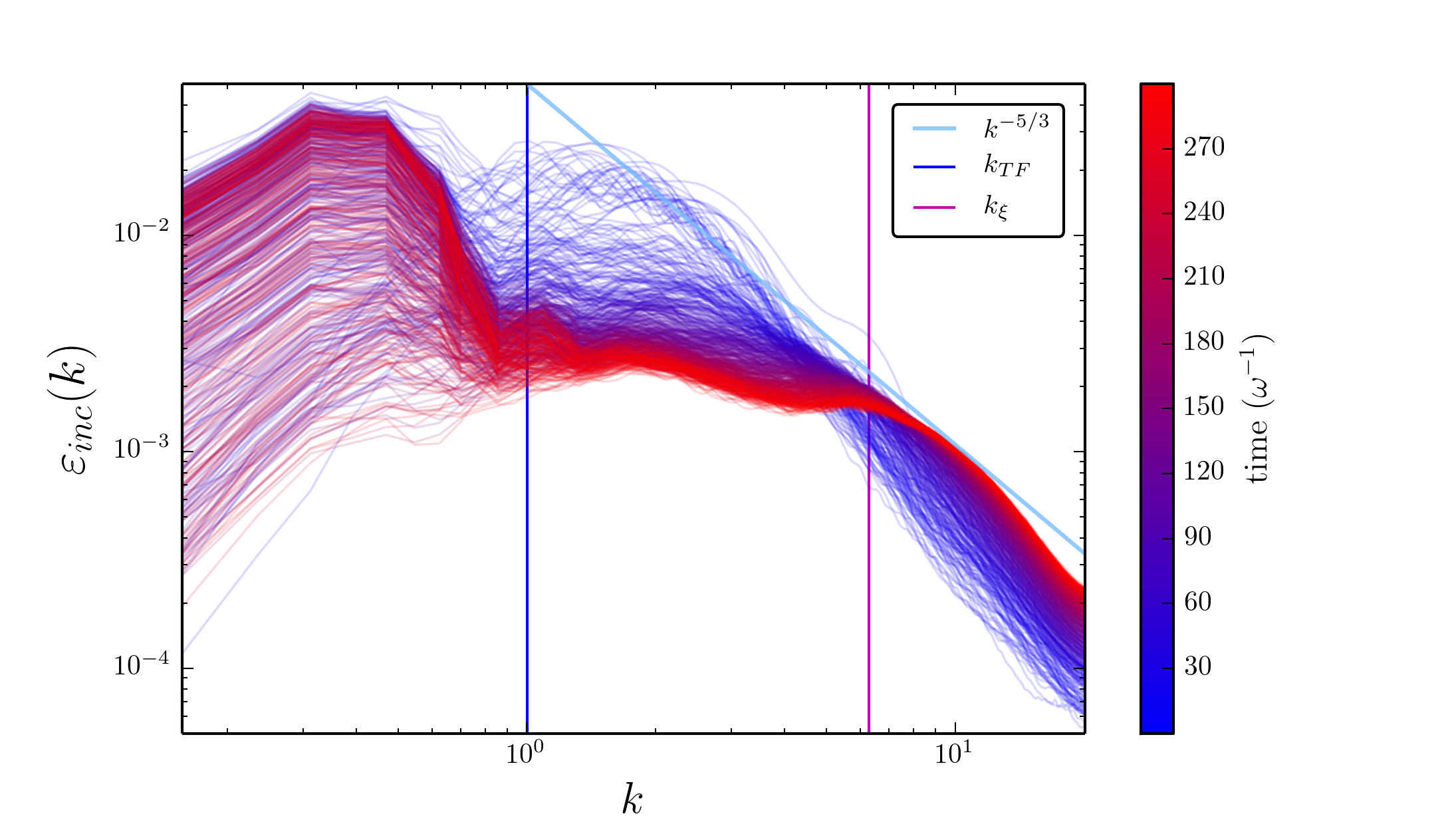}
\end{center}
\caption{(Color online) Time evolution of the kinetic incompressible energy spectrum of an initial vortex ring. See Fig. \ref{ene-cruzados} for explanation of the straight lines.}
\label{ene-ring}
\end{figure*}

\begin{figure*}[ht]
\begin{center}
\includegraphics[width=1.0\textwidth,keepaspectratio]{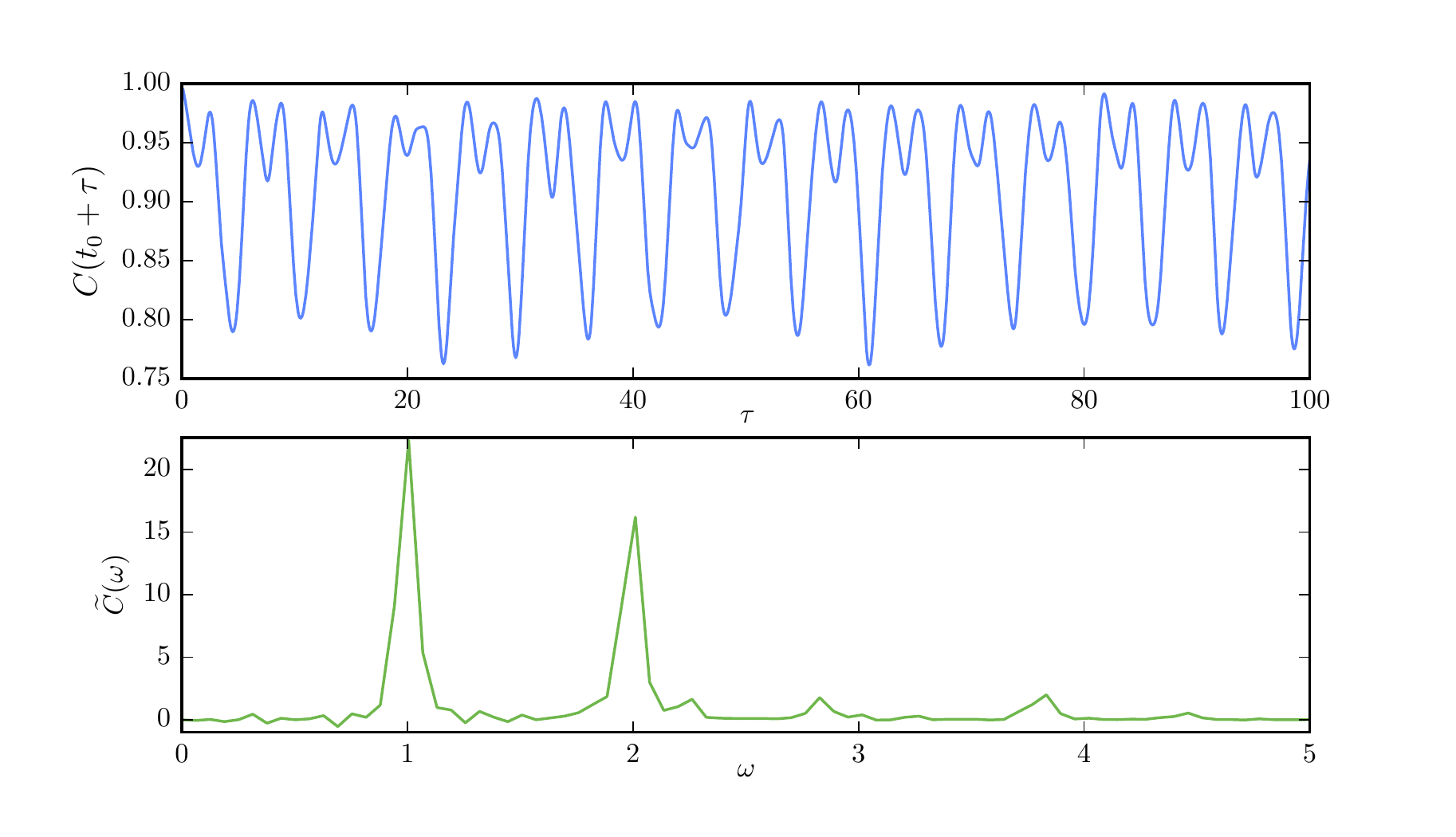}
\end{center}
\caption{(Color online) Overlap $C(t_0 + \tau)$ vs $\tau$, see Eq.(\ref{overtime}), for $t_0 =200$ in the upper panel, and its Fourier transform $\tilde C(\omega)$ vs $\omega$  in the lower one, for an off-center vortex ring excitation.}
\label{overlapRingNC}
\end{figure*}

This case corresponds to the evolution of a single vortex ring, shown in Fig. \ref{ring}, initially placed at $(x_0,y_0,z_0) = (1.8,0.51,-1.57)$ and $Q=-1$, with  
\begin{equation}
\theta(\vec{r}) = \arctan\left( \frac{z - z_0}{\sqrt{(x-x_0)^2+(y-y_0)^2 } - 1}\right)
\end{equation}
We find that, due to its circulation, the ring starts to move to one of the poles of the condensate, see Fig. \ref{ring}, expanding at the surface of the cloud ($t = 10.0$), returning to the inside but then breaking into internal excitations and what appears to be a vortex at the edge of the condensate ($t = 20.0$ and $t = 30.0$). Referring to Fig. \ref{fidelity0All}, we find that this excitation does not loose completely its time-reversal invariance, but again, once it enters the stationary state remains there, see Fig. \ref{fidelity200All_v1}. As we see below, the stationary state is composed of oscillations of the whole cloud, some collective excitations and certainly with the presence of phonons. This can be concluded from the incompressible kinetic energy spectra and the overlap function, eq. (\ref{overtime}), see Figs. \ref{ene-ring} and \ref{overlapRingNC}. This stationary state appears to be close to the previous one of two colliding anti vortices, with a strong presence of phonons, yet with collective modes playing a role. 

\subsection{Vortex of charge $Q = 2$}

\begin{figure*}[ht]
\begin{center}
\includegraphics[width=1.0\textwidth]{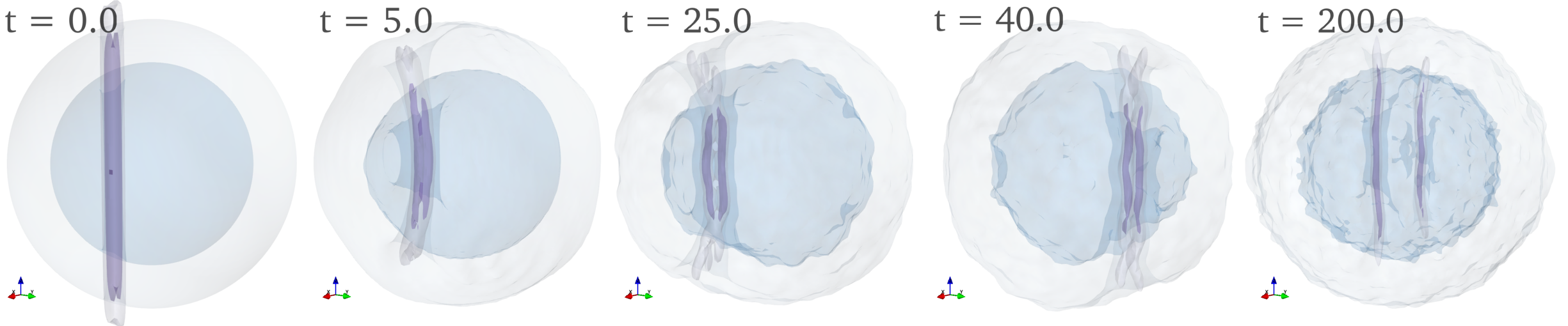}
\end{center}
\caption{(Color online) Snaphots of the magnitude of the velocity field for the evolution of an initial Q = 2 perturbation.}
\label{Q2}
\end{figure*}

\begin{figure*}[ht]
\begin{center}
\includegraphics[width=1.0\textwidth,keepaspectratio]{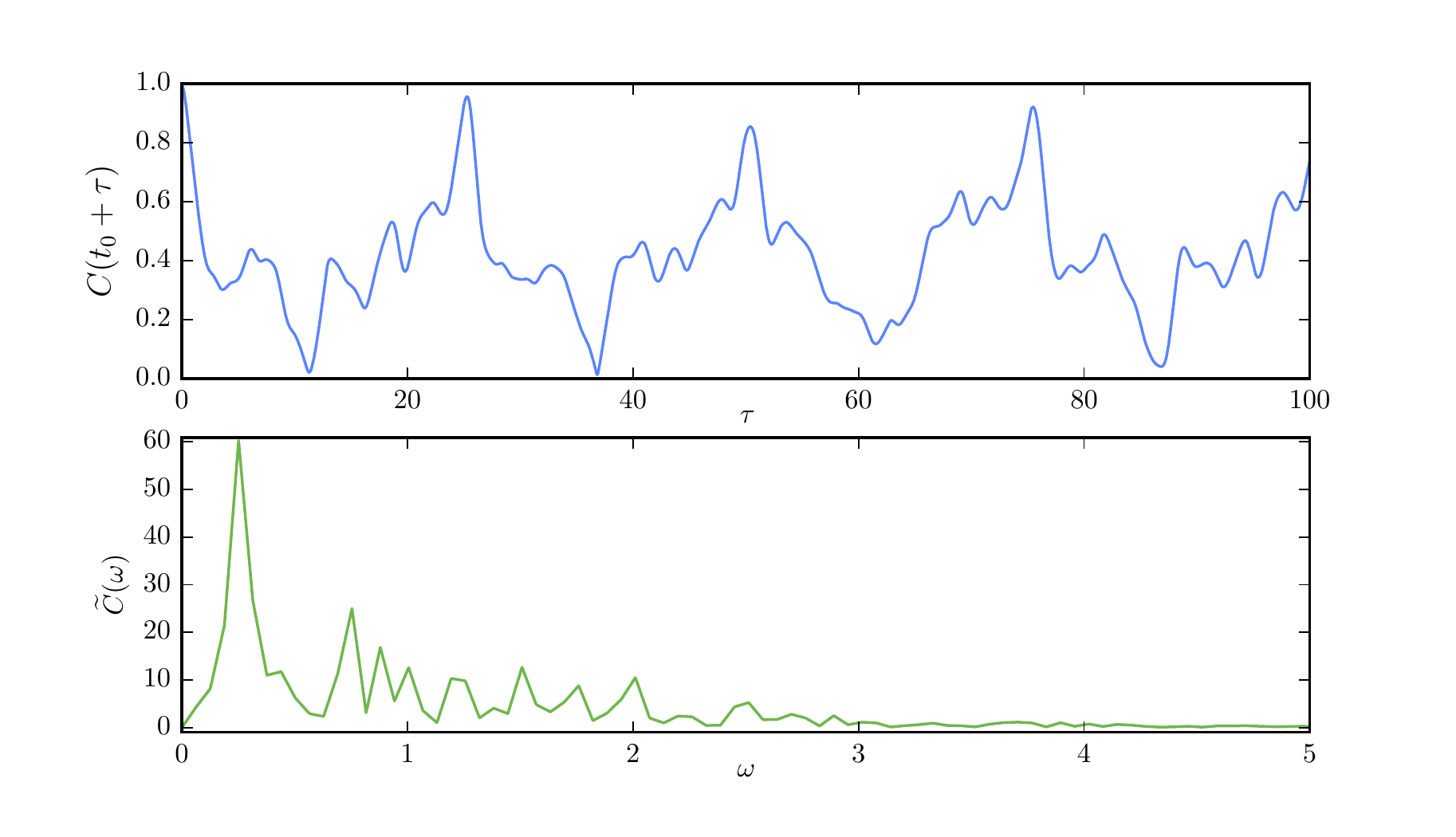}
\end{center}
\caption{(Color online) Overlap $C(t_0 + \tau)$ vs $\tau$, see Eq.(\ref{overtime}), for $t_0 =200$ in the upper panel, and its Fourier transform $\tilde C(\omega)$ vs $\omega$  in the lower one, for a vortex of charge $Q = 2$.}
\label{overlapQ2}
\end{figure*}

We now analyze the case in which the initial excitation is one that corresponds a single line vortex with topological charge $Q = 2$, parallel to $z$, at $x = 2.0$ and $y = 0$, see Fig. \ref{Q2}. 
As it is known,\cite{Castin} vortices of charge $Q > 1$ are unstable and decay into two vortices of the same charge $Q = 1$. This situation is exemplified in the snapshots of times $t =  5.0$, 25.0 and  40.0 of Fig. \ref{Q2}. It is very interesting to see that right after its creation, the vortex separates into two braid-entangled vortices, then these unwind until  two stable parallel vortices are formed. The evolution dephases after 50 units of time, see Fig. \ref{fidelity0All}, yet, as with the other cases, it enters a stationary state. This stationary state, however, is the least stable, as can be seen from Fig. \ref{fidelity200All_v1}, where the fidelity appears to be lossing about 5 \% of its value, and thus, for longer times the system may migrate to a different stationary state. This remains to be verified. In any case, judging from Figs. \ref{Q2} and \ref{overlapQ2}, the stationary state shows mainly vortex excitations with collective excitations of the cloud, similar to the case of crossed vortices. The large peak in the overlap spectrum in Fig. \ref{overlapQ2} corresponds the fast orbiting of the two vortices around each other.

\begin{figure*}
\begin{center}
\includegraphics[width=1.0\textwidth]{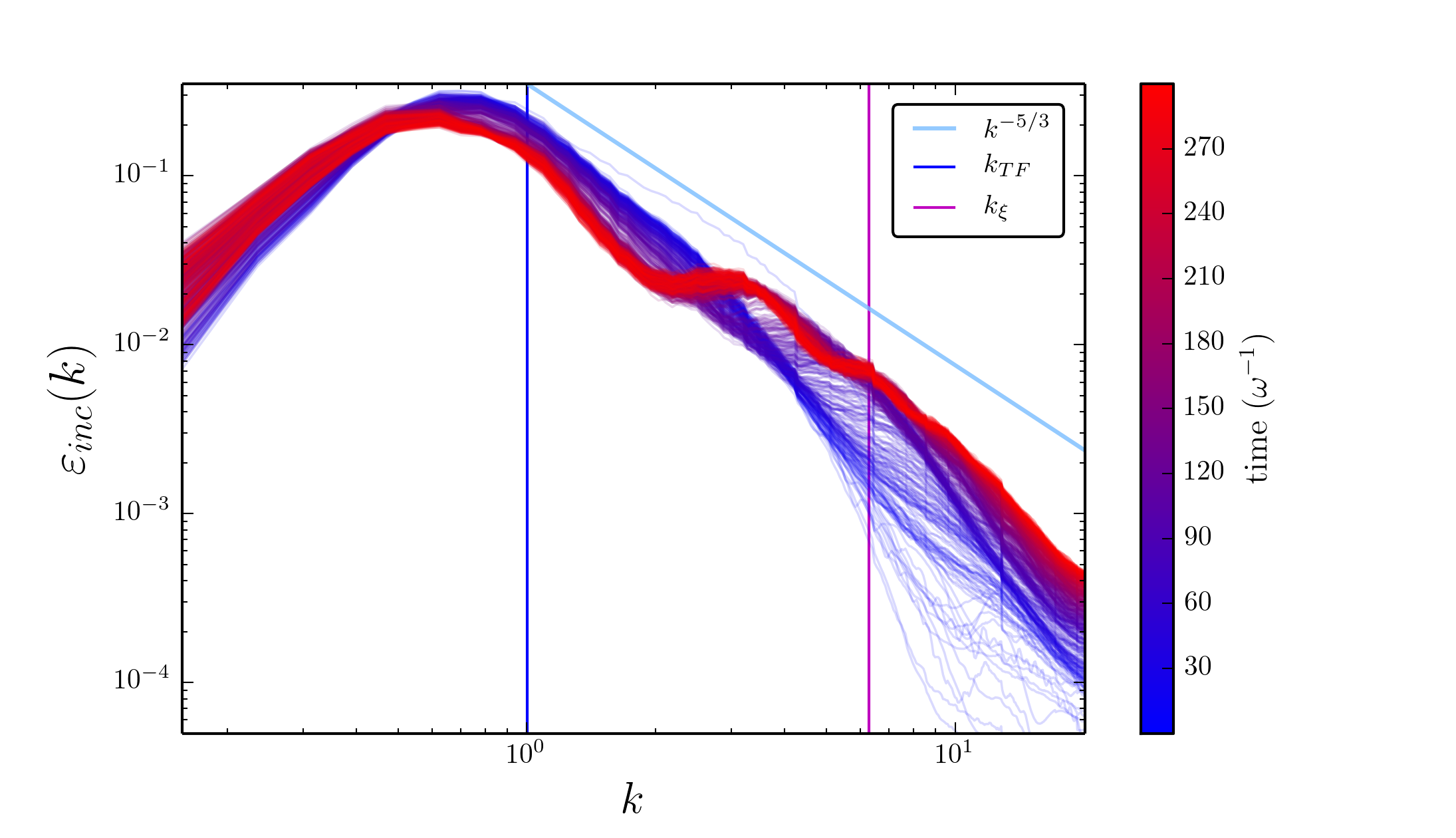}
\end{center}
\caption{(Color online) Time evolution of the incompressible kinetic energy spectrum of an initial Q = 2 perturbation. See Fig. \ref{ene-cruzados} for explanation of lines.}
\label{ene-Q2}
\end{figure*}

\subsection{A tangle of 4 vortices}

\begin{figure*}[ht]
\begin{center}
\includegraphics[width=1.0\textwidth]{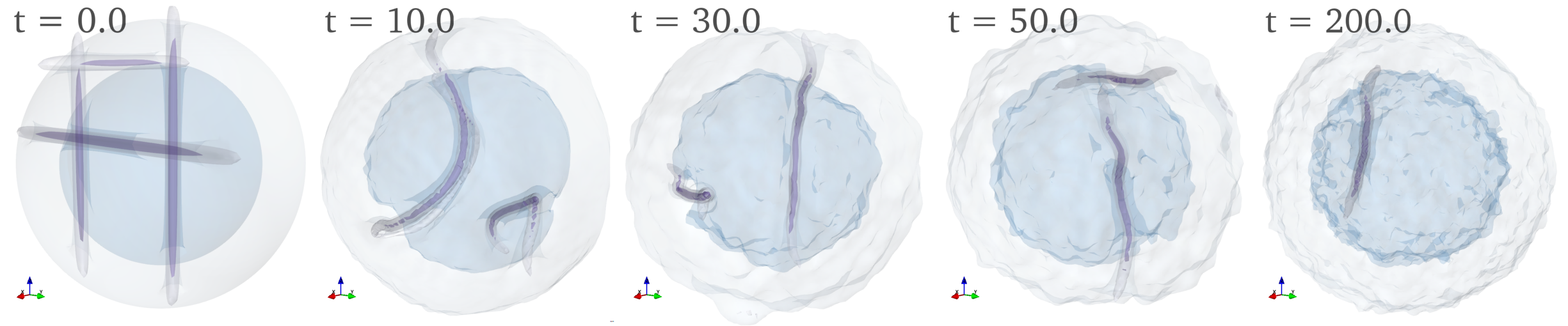}
\end{center}
\caption{(Color online) Snaphots of the magnitude of the velocity field for the evolution a tangle of 4 vortices, two parallel anti vortices and two crossed ones.}
\label{VT}
\end{figure*}

\begin{figure*}[ht]
\begin{center}
\includegraphics[width=1.0\textwidth]{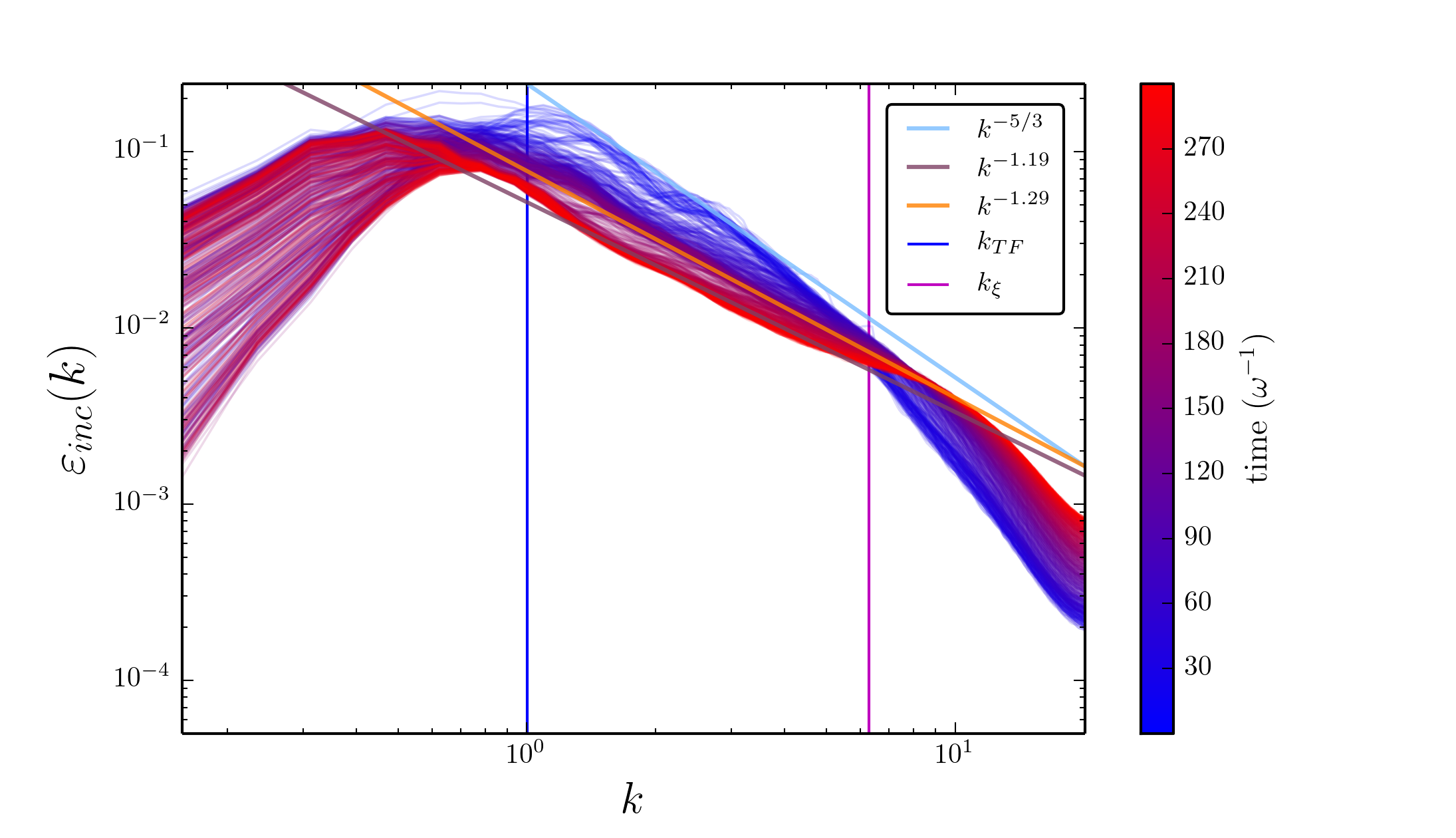}
\end{center}
\caption{(Color online) Time evolution of the incompressible kinetic energy spectrum of a tangle of 4 line vortices. See text for explanation of the three slope fits.}
\label{ene-Tangle}
\end{figure*}

\begin{figure*}[ht]
\begin{center}
\includegraphics[width=1.0\textwidth,keepaspectratio]{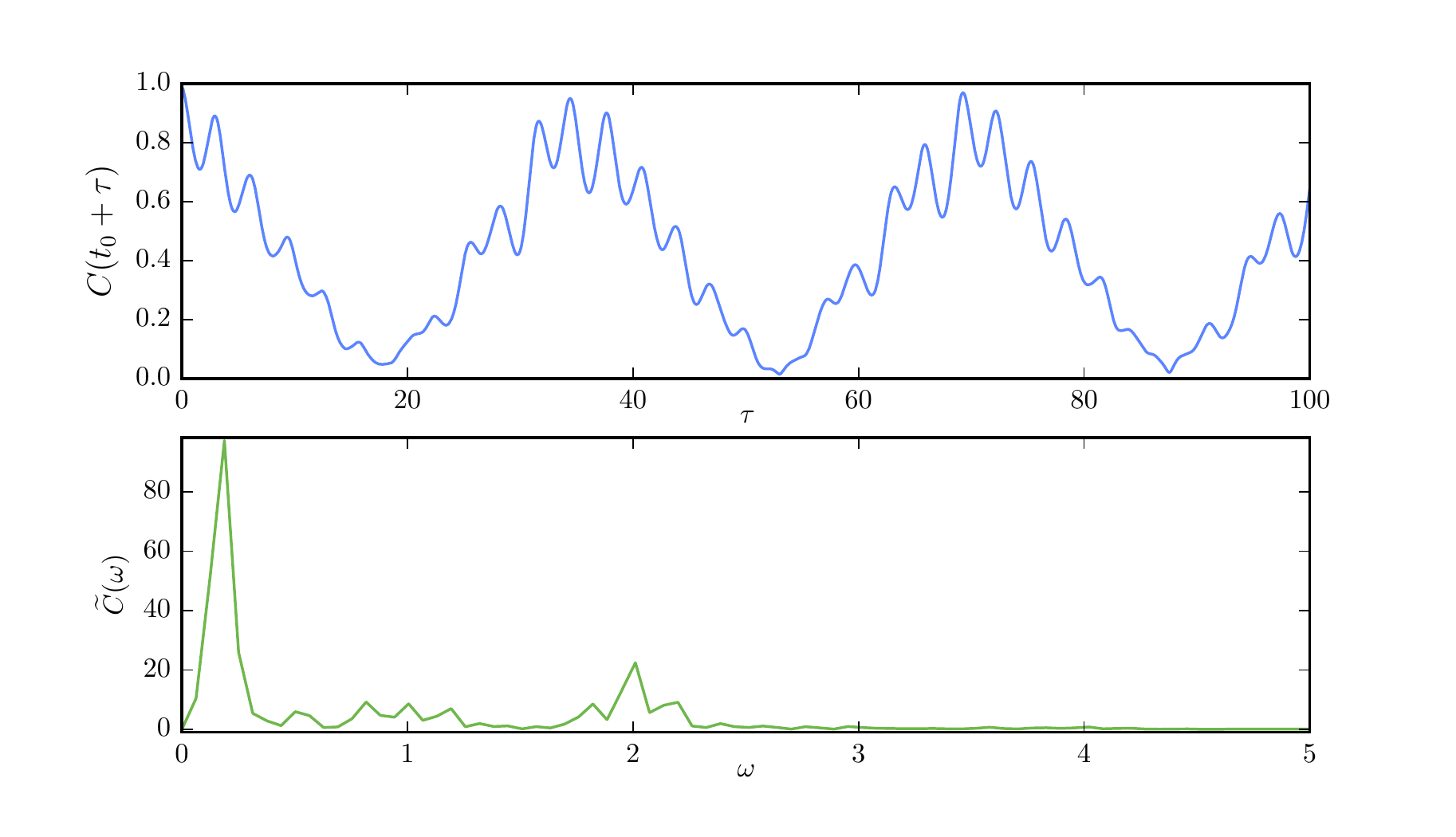}
\end{center}
\caption{(Color online) Overlap $C(t_0 + \tau)$ vs $\tau$, see Eq.(\ref{overtime}), for $t_0 =200$ in the upper panel, and its Fourier transform $\tilde C(\omega)$ vs $\omega$ in the lower one, for a tangle of 4 vortices.}
\label{overlapRND4}
\end{figure*}

Here we study a ``tangle" of 4 vortices, two anti-vortices parallel to the $z$-axis at random positions $(x_1,y_1) = (1.20,-2.99)$ and $(x_2,y_2) = (-0.64,0.05)$, with $Q_1 = +1$ and $Q_2 = -1$; and two cross vortices, one parallel to the $x$-axis at $(y_3,z_3) = (-1.86,3.60)$ with $Q_3 = -1$, and the other parallel to the $y$-axis at $(x_4,z_4) = (2.57,1.05)$ with $Q_4 = +1$, see Fig. \ref{VT}. This would be the simplest tangle of vortices that may lead to turbulence, as originally suggested by Tsubota et al.\cite{Kobayashi_Tsubota_PRA_2007} and also already treated by White et al.\cite{AC_White_JOP_CS_2011} The evolution of this case is very harsh, similarly to the case of the colliding anti-vortices; here we see that by $t = 10$ there have been several reconnections and there remain two vortices only. Then the system enters a very agitated flow, perhaps turbulent for times $t \approx 30$ to $t \approx 50$. Unfortunately, the dephasing time is about 50 units of time, and therefore the true evolution is lost for longer times. Nevertheless, after this agitated state, one of the vortices disappears, and the system enters a stable stationary state. These stages can also be observed in the evolution of the incompressible kinetic energy spectrum, Fig. \ref{ene-Tangle}: the slope changes from nearly $-5/3$ to $-1.29$ to settle about $-1.19$. This last stage corresponds to the stationary state which, from observing the overlap $C(t_0 + \tau)$ and its Fourier transform in Fig. \ref{overlapRND4}, appears to be composed of a fast vortex, the largest peak, with collective excitations and motion of the whole cloud, but with very few phonons. 

\section{Final Remarks}

In this article we have analyzed the evolution of several representative vortex states in an otherwise equilibrium Bose-Einstein condensate, using Gross-Pitaevskii e\-qua\-tion as a model for a gas at very low temperatures. Our goal is to add to the understanding of evolution of complicated states in the search for different way to approach a quantum turbulent state. 
Since the dynamics observed show different evolutions in wide time scales, and these types of studies can only be realized numerically, one must elucidate the role of rounding errors in determining the fate of the evolution. We have found that, depending on the initial state, the evolved state indeed looses its time-reversal invariance but it does so at different late times. Moreover, the system reaches a stationary state in which time-reversal invariance is robust. 

Referring to the particular initial states considered in the present article, we can say that they are widely divided in two types: (I) those that at the stationary state have one vortex or an array of vortices with collective excitations of different sizes, and (II) one with an agitated state that appears dominated by Bogoliubov phonons. In case (I) one can include the initial states with two crossed vortices (3.1), the vortex of charge two (3.4) and the tangle of 4 vortices (3.5); in case (II), we find the two vortex-antivortex system (3.2) and the off-center ring vortex (3.3). This wide classification is based both on the spectra of the incompressible kinetic energy and in the overlap function in the stationary state, eq. (\ref{overtime}). In the relevant wavenumber range, from $k_{TF}$ to $k_\xi$, one corresponding to the size of the cloud, the other to the vortices cores, the kinetic energy spectra in case (I) appears to be close to the Kolmogorov law $k^{-5/3}$, while in (II) the spectra is very different even with positive power law, but that matches very well the Bogoliubov phonon dispersion relation. 

While finding a similarity with Kolmogorov law $k^{-5/3}$ does not necessarily implies the presence of turbulence, it nevertheless suggests that there is a transfer or interchange of energy between excitations of different sizes, without a characteristic length scale, namely, with a scaling algebraic law. We recall that Kolmogorov derivation of his law\cite{Kolmogorov,Landau-FluidMechanics} involves two parts, one in which a scaling law is assumed for the velocity flow, and a second one, the elucidation of the energy transfer rate between excitations of different sizes. The exponent $-5/3$ is a consequence of the scaling, independent of the energy transfer mechanism.  Thus, we hypothesize that the observed similarity with Kolmogorov law in some cases, essentially the same for the two crossed vortices, indicates the presence of energy transfer among excitations at different length scales: typically a single vortex, that while rotates and orbits around the cloud, interacts with collective cloud excitations. This gives rise clearly to a complicated, agitated flow. Whether this may be considered ``turbulence" remains to be fully elucidated. It is, therefore, the more important to better understand why colliding antivortices and ring vortices appear to be destroyed, with the accompanying creation of acoustic phonons, yielding also a chaotic state, but with a completely spectral law.

\begin{acknowledgements}
We thank support from grant DGAPA-UNAM IN107014. RZZ and OAA acknowledge support from CONACYT-Mexico.
\end{acknowledgements}

\pagebreak

\end{document}